\documentstyle[prl,aps,epsf,rotate,preprint]{revtex}

\begin{document}
\draft

\title{Direct Instantons in QCD Nucleon Sum Rules}

\author{Hilmar Forkel and Manoj K. Banerjee}
\address{Department of Physics, University of Maryland, College Park,
Maryland 20742-4111 (U.S.A.)}

\maketitle

\vskip 2cm

\centerline{February 1993}

\vskip .5cm

\centerline{hep-ph/9309232}

\begin{abstract}

We study the role of direct (i.e. small-scale) instantons in QCD
correlation functions for the nucleon. They generate sizeable,
nonperturbative corrections to the conventional operator product
expansion, which improve the quality of both QCD nucleon sum rules
and cure the long-standing stability problem, in particular, of the
chirally odd sum-rule.

\end{abstract}

\narrowtext

QCD sum rules, introduced by Shifman, Vainshtein and Zakharov (SVZ)
\cite{SVZ79}, provide a systematic, nonperturbative framework for the
calculation of hadron properties. They have been intensely studied and
applied over the last decade and produced the most exhaustive
model-independent analysis of hadron properties to date.

The sum-rule approach is based on the comparison of two ``dual''
descriptions for correlation functions of hadronic currents, in
terms of quarks and gluons on the left-hand side (LHS) and in terms
of hadrons on the right-hand side (RHS). The RHS uses a simple
parametrization of the spectral function in terms of the hadron
parameters, such as mass, overlap with the source current and continuum
threshold. The QCD calculation on the LHS employs a non-perturbative
operator-product expansion (OPE). Long-distance bulk properties of
the physical vacuum are efficiently parametrized in terms of the
vacuum expectation values of composite quark-gluon operators
(``condensates''), which are independent of the hadron considered.
The short-distance physics is contained in the perturbatively
calculated Wilson coefficients. The inverse renormalization scale
of the operators, $\mu^{-1}$, serves as the dividing line separating
long and short distances.

Both sides of the sum rules are then Borel-transformed, and the hadron
parameters are determined by fitting the two sides in the fiducial
region, i.e. in the range of Borel mass values in which both
descriptions of the correlator are expected to be adequate. The
quality of this fit is the only intrinsic criterion for the accuracy
and reliability of the sum rules. If it is met sufficiently well, the
resulting hadron parameters will be approximately independent of the
Borel mass in the fiducial region. The nucleon sum rules, however, do
not show such a stability plateau, despite many improvements over the
last decade \cite{Ioffe81}. It seems that some relevant physics in the
fiducial region (around 1 GeV) is missing in the OPE. In this letter
we suggest that small-size instantons \cite{fn1}, termed ``direct''
by SVZ, provide the dominant part of this physics.

Instantons\cite{BPST75} are classical solutions of the euclidean
Yang-Mills equation. Due to the infrared complexities of QCD, their
quantum properties and vacuum distribution cannot yet be derived
from first principles. A consistent picture of their importance and
bulk features has been established, however, by extensive
phenomenological \cite{Shu82,instliq}, analytical \cite{DiaPe84}
and numerical studies (in the instanton liquid model \cite{instliq}
and on the lattice, e.g. in ref. \cite{ChuHua92}). They indicate,
in particular, that the average instanton size $\rho_c$ in the vacuum
is considerably smaller than the average separation $R$ between
instantons \cite{instliq}:
\begin{eqnarray}
\rho_c \simeq \frac{1}{3} \,{\rm fm}, \quad \quad R \simeq 1 \,
{\rm fm}. \label{bulkprop}
\end{eqnarray}
Most of the contributions of these rather small instantons to the
correlation function are ignored in the conventional OPE: As the
scale of these fields is smaller than the inverse renormalization
scale (taken around $\mu \simeq 0.5 {\rm GeV}$), they would
contribute to the Wilson coefficients, but do not show up in the
perturbative evaluation.

The aim of our paper is to calculate the leading instanton
contributions to the nucleon correlator, using an instanton
size-distribution \cite{instliq} in accordance with the bulk
features (\ref{bulkprop}), and to study their effects in the
nucleon sum rules. We do not use the detailed assumptions of
the above-mentioned model calculations.

The instanton contributions are mediated mainly by the quark
zero-modes \cite{tHoo76} in the instanton background field.
Due to their particular chiral and color properties, the
magnitude of instanton effects is channel dependent. Even
if they can {\it e.g.} be safely neglected in the vector
and axial-vector channel, they play a dominant role in the
pseudoscalar sum rules  \cite{Shu83} (and more generally
in the spin-0 channel), where the conventional OPE was known
to fail \cite{SVZ79}.

Our expectation of sizeable instanton effects in the nucleon
sum rules is mainly based on parallels with the pseudoscalar
sector. The interpolating fields (see below) in the nucleon
correlator contain spin-0 diquarks, which receive zero-mode
contributions of the same order as the pseudoscalars.
Furthermore, the magnitude of the nucleon correlator at
distances around 1 - 2 fm is much larger than the perturbative
contribution \cite{Shu92}, which is reminiscent of the strongly
attractive correlations due to instantons in the pseudoscalar
channel.

Recently, Dorokhov and Kochelev \cite{DoKo90} made a first
attempt to calculate instanton contributions to the nucelon
sum rule. However, as we will outline in the course of this
paper, we do not agree with their results and many of their
conclusions.

The QCD nucleon sum-rules are based on the correlation function
\begin{eqnarray}
{\rm i} \int \, d^4 x \, {\rm e}^{{\rm i} qx} <0| \,T \,
\eta (x) \, \overline{\eta} (0)\,|0> \,\, = \,\, \rlap/{q}\,
\Pi_q (q^2) + \Pi_m (q^2), \label{corr}
\end{eqnarray}
evaluated at intermediate space-like momentum transfer, $Q^2
\equiv -q^2 \simeq 1 {\rm GeV}$. We will use the most general
interpolating field  of the nucleon (with minimal mass dimension):
\mediumtext
\begin{eqnarray}
\eta (x) = \epsilon_{a b c} \left\{ \,[u_a^T (x)C d_b(x)]
\gamma_5 u_c(x) + t \, [u_a^T(x) C \gamma_5 d_b(x)]  u_c(x)
\, \right\}.
\end{eqnarray}
\narrowtext
Here, $u,d$ denote the up and down quarks, $C$ is the
charge-conjugation Dirac matrix, and the real coefficient
$t$ specifies the linear combination of the currents
containing scalar and pseudoscalar diquarks. Ioffe's
current \cite{Ioffe81} corresponds to $t= -1$.

We intend to evaluate the dominant instanton contributions
to the correlator eq. (\ref{corr}) at distances relevant
for the sum rules, $q^{-1} \simeq  0.2\, {\rm fm}$ \cite{SVZ79}.
As these distances are small compared to the average separation
$R\simeq 1 \, {\rm fm}$ between instantons, contributions from
a single instanton should dominate multi-instanton effects.
The small average size of the instantons further implies a
sufficiently small gauge coupling $g(\Lambda \rho_c)$, which
allows us to calculate the correlator in semiclassical
approximation.

To this end, we evaluate eq. (\ref{corr}) with the help of
the quark propagator in an instanton background field
\cite{Shu82}, which is dominated by the zero-mode contribution
\begin{eqnarray}
S_0^{\pm}(x,y)  = \frac {\psi_0^{\pm}(x) \psi_0^{\pm \,
\dagger} (y) }{m^*(\rho) } + O( (\rho m^*)^{-1} ).
\label{zeroprop}
\end{eqnarray}
The superscript $\pm$ refers to an instanton/anti-instanton
of fixed size $\rho$ and position $x_0$. The zero-mode
solutions $\psi_0^{\pm}$ have been derived by 't Hooft in
ref. \cite{tHoo76}. Due to interactions with QCD vacuum
fields, the quarks acquire an effective mass, $m^*(\rho) =
m_q -\frac{2}{3} \pi^2 \rho^2 <0| \,\overline{q} q \,|0>$,
which has been given by SVZ \cite{SVZ80a}. (In the following,
we will neglect the small current quark masses $m_q$.)

Due to the symmetry properties of the zero-mode solution,
maximally two of the three quarks created by the interpolating
field propagate in zero-mode states.  We approximate the
propagation of the third quark in the continuum states
\cite{DiaPe84} by the standard quark propagator used in
the OPE and obtain
\widetext
\begin{eqnarray}
<0| \,T \, \eta (x) \, \overline{\eta} (0)\,|0> \,\, &=& \,
\frac{1}{\pi^4} \left( \frac{c_1}{\pi^2}
\frac{\rlap/{x}}{x^4} + \frac{2 i  c_2}{N_c} <0|
\,\overline{q} q \,|0> \right) \nonumber \\ & \times &
\int d \rho \, n(\rho)    \frac{\rho^4}{m^{* 2}(\rho)}
\int d^4 x_0 \frac{1}{((x-x_0)^2 + \rho^2)^3}
\frac{1}{(x_0^2 + \rho^2)^3}, \label{xcorr}
\end{eqnarray}
\narrowtext
\noindent where we define
\begin{eqnarray}
 c_1 = 6(t^2 - 1), \quad c_2 = \frac{1}{8} \,
\left[ 13(t^2 + 1) + 10 t \, \right].
\end{eqnarray}

Due to translational and gauge invariance, the
integration over all collective coordinates of the
instanton except the scale $\rho$ is straightforward.
The broken scale invariance, however, gives rise to
a $\rho$-dependent weighting factor, the instanton
size-density $n(\rho)$. As motivated above, we will
use the simple form proposed in ref. \cite{Shu82},
\begin{eqnarray}
n(\rho) = n_c \, \delta (\rho - \rho_c), \label{sizedistr}
\end{eqnarray}
which incorporates the already described features and
scales of the instanton distribution and the quite
sharply peaked (half-width $\simeq 0.1 {\rm fm}$),
almost gaussian result of Monte-Carlo simulations
\cite{ShuVer90}.

The Fourier transform of (\ref{xcorr}) contains the
instanton contribution to the invariant amplitudes
$\Pi_q (q^2)$ and $\Pi_m (q^2)$ in (\ref{corr}). Their
subsequent Borel transform, required in the sum rules,
leads to quite complicated integrals \cite{FoBan92}
which we evaluate numerically, but do not write down
explicitly here. Instead we present a saddle-point
approximation, which is more transparent and allows
for a direct comparison with the work of \cite{DoKo90}:
\widetext
\begin{eqnarray}
\hat{ \Pi}_q (M_B^2) = \frac{c_1}{64 \sqrt{\pi} \pi}
\frac{n_c}{m^{* 2} \rho_c^2}  \left[ \frac{64}{10
\sqrt{\pi}} (\frac{1}{z^2} - \frac{24}{7}
\frac{1}{z^4}) + ( z^2 +4 +9 \frac{1}{z^2})
\frac{e^{-z^2}}{z} + 3 \sqrt{\pi} ( \frac{1}{z^2} +
\frac{15}{2z^4}) \, {\rm erfc}(z) \right] \label{piq}
\end{eqnarray}
\narrowtext
\noindent and
\begin{eqnarray}
\hat{ \Pi}_m (M_B^2) = \frac{- \sqrt{\pi} c_2}{4 N_c }
\frac{<\overline{q} q > n_c }{{m^*}^2} (z^3 -
\frac{3}{2} z) e^{- z^2}, \label{pim}
\end{eqnarray}
where $M_B$ is the Borel mass, ${\rm erfc}(z)$ is the
complementary error function \cite{Abram72} and we
define $z = M_B \rho_c$. Note that the instanton
contribution to $\hat{ \Pi}_q$ vanishes for the Ioffe
current (i.e. $c_1 = 0$).

We eliminate the $n_c$ dependence of (\ref{piq}) and
(\ref{pim}) with the self-consistency relation \cite{CDG}:
\begin{eqnarray}
<\overline{q} q > = -2 \int d \rho \,
\frac{n(\rho)}{m^*(\rho)} = -2\frac{n_c}{m^*(\rho_c)} .
\label{selfcon}
\end{eqnarray}
While the assumption of a completely zero-mode-induced
quark condensate is rather strong, our final results
will not depend on it, because the numerical value
obtained from (\ref{selfcon}) is almost exactly equal
to the phenomenologically estimated value.

Before combining eqs. (\ref{piq}) and (\ref{pim}) with
the standard OPE terms of the nucleon sum rules
\cite{Ioffe81}, we have to address the issue of double
counting instanton physics. The conventional OPE
already accounts for {\it all} nonpertubative vacuum
fields ({\it including} instantons) with scales {\it
below} the renormalization scale $\mu$ in the condensates,
whereas the whole $M$-dependence, logarithms and powers
of $M^{-2}$, comes from the perturbatively calculated
Wilson coefficients.

The $M$-dependence of the instanton contributions,
however, is  exponential in all terms of $\hat{\Pi}_m$
and in all but two terms of $\hat{\Pi}_q$. Hence these
terms constitute new, nonpertubative contributions to
the Wilson coefficients, originating from small-instanton
physics. $\hat{ \Pi}_q$, however, contains also two
power terms \cite{fn3}, which represent instanton
contributions to the condensates of dimension 6
(four-quark condensates) and 8 \cite{fn4}. In this
letter we will restrict the conventional OPE to operators
of dimension $\le 6$, so that double counting has to be
prevented only for the four-quark condensates.

The standard  sum rule practice is to approximate the
four-quark condensates by factorizing them into two-quark
condensates, and it is known that this approximation
doesn't always work well for the instanton contribution.
Indeed, the four-quark condensate term in eq.
(\ref{piq}) is considerably larger than the corresponding
factorized OPE condensate, which we will therefore
omit completely.

The LHS of the sum rules can now be obtained by adding
the invariant amplitudes, (\ref{piq}) and (\ref{pim}),
to their standard OPE counterpart \cite{Ioffe81} with
the corrected four-quark condensate. On the RHS, these
amplitudes are written in a (Borel-transformed) Lehmann
representation with the usual pole-continuum
approximation for the spectral functions \cite{Ioffe81}.
Equating both sides and transferring the continuum
contribution to the left, we obtain the two sum rules
\widetext
\begin{eqnarray}
 2 c_3 M^6 A_2 (M,W) + 2 c_3 \pi^2 M^2 <
\frac{\alpha}{\pi} G_{\mu \nu} G^{\mu \nu} >  A_0 (M,W)
+ \frac{1}{3} c_5 (4 \pi)^4 <\overline{q} q >^2  +
\frac{96}{5} c_1 \rho_c^{-6} \left(1 - \frac{24}{7}
z^{-2} \right)\nonumber \\ + 3  c_1\sqrt{\pi} \rho_c^{-6}
\left[ z^3 + 4 z + \frac{9}{z} + 3 \sqrt{\pi} \left(1 +
\frac{15}{2} z^{-2} \right) e^{z^2} {\rm erfc}(z) \right]
e^{-z^2} = \tilde{\lambda}_N^2 e^{-m^2/M^2} \label{qsumr}
\end{eqnarray}
and
\begin{eqnarray}
 -(4\pi)^2 <\overline{q} q > \left[ c_4 M^4 A_1 (M,W) -
\frac{c_1}{8} m_0^2 M^2  A_0 (M,W) +  c_2 \sqrt{\pi}
\rho_c^{-4} z^3 \left( z^2 - \frac{3}{2} \right) e^{-z^2}
\right] = m \tilde{\lambda}_N^2  e^{-m^2/M^2}, \label{msumr}
\end{eqnarray}
\narrowtext
\noindent where the
\begin{eqnarray}
A_n(M,W) = 1 - e^{-W^2/M^2} \left[ 1 + \sum_{m=1}^n
\frac{1}{m} \left(\frac{W^2}{M^2} \right)^m \right]
\end{eqnarray}
contain the continuum contributions and
\begin{eqnarray}
c_3 = \frac{1}{8} \left[ 5 ( t^2 + 1) + 2 t \right],
\quad c_4 = \frac{1}{4} \left[ 7 t^2 - 2 t - 5  \right],
\quad c_5 = \frac{1}{8} (t - 1)^2. \nonumber
\end{eqnarray}
Let us now compare our results with those of ref.
\cite{DoKo90}, where semi-classical instanton contributions
to the nucleon sum rules are also evaluated. We find four
additional terms in $\hat{\Pi}_q$ and one in $\hat{\Pi}_m$,
all with exponential Borel-mass dependence. These terms,
which apparently have been missed in \cite{DoKo90}, are in
no way subleading and their omission has drastic consequences:
1.) The spectral function of $\hat{\Pi}_q$ becomes negative
in the whole fiducial domain, thus violating its very general
positivity bound \cite{BjoDre}. 2.) For the same reason,
the (usually more reliable) $\hat{\Pi}_q$ sum rule becomes
unstable and has to be abandoned. Half of the physical
information and the valuable consistency check between the
two sum rules is lost. 3.) A numerically important part of
$\hat{\Pi}_m$ is missing and the relation between nucleon
mass and quark condensate, usually manifest in the ``Ioffe
formula'' \cite{Ioffe81}, is obscured.

Furthermore, the authors of \cite{DoKo90} used the (only
qualitatively reliable) saddle-point approximation to evaluate
the sum rules, which induces substantial errors ($\sim$ 30
\%) in the nucleon parameters. (As already mentioned, our
results are based on the exact, numerical evaluation of the
instanton contribution.) Finally, double-counting of the
four-quark condensate is not addressed in \cite{DoKo90}
and the 4- and 5-dimensional condensates are neglected.

The free parameters to be determined in the sum rules are
the continuum threshold $W$, the nucleon mass $m$ and
$\lambda_N$, the overlap of the wave packet created by
$\eta$ with the nucleon $(<0| \,\eta (0) \,|N_{\alpha}(k)>
= \lambda_N \, u_{\alpha}(k), \, \tilde{\lambda}_N \equiv
(4 \pi)^2 \lambda_N)$. We obtain their values by minimizing
the difference, defined by the measure $\delta$ of ref.
\cite{Ioffe81}, between the two sides of the sum rules
(\ref{qsumr}) and (\ref{msumr}) in the fiducial Borel mass
domain $ 0.8 {\rm GeV} \le M \le 1.2 {\rm GeV} $. In this
letter we choose $t = -1.1 $ \cite{fn7}, corresponding
to the optimal current of Espriu et al. \cite{Ioffe81}
and close to the Ioffe current. For {\it any} choice of
the current, however, the instanton contribution will be
substantial in at least one sum rule.

Figure 1a shows the OPE contribution, the instanton
contribution and their sum, the complete LHS, in comparison
with the pole term $\tilde{\lambda}_N^2 e^{-m^2/M^2}$ for
the $\rlap/{q}$ sum rule (\ref{qsumr}). Figure 1b shows
the same curves, divided by the nucleon mass $m$, for the
sum rule (\ref{msumr}). As expected, the instanton
contributions to the $\rlap/{q}$ sum-rule are rather small
for our current, but still improve the fit quality ($\delta
= 1.2 \%$). The nucleon parameters become $m =0.88 {\rm GeV}$,
$W = 1.43 {\rm GeV}$ and $\tilde{\lambda}_N^2 = 2.90
{\rm GeV}^{6}$.

In the traditionally less stable $m$ sum-rule, however, the
instanton contributions become comparable to the OPE
contributions and improve the fit quality dramatically.
Their $M$ behaviour balances the OPE terms over the full
Borel-mass interval to bring their sum very close ($\delta
= 0.03 \%$) to the pole contribution. The nucleon parameters
become $m = 0.90 {\rm GeV}$, $W = 1.60 {\rm GeV}$ and
$\tilde{\lambda}_N^2 = 5.68 {\rm GeV}^{6}$.

The different magnitude of the instanton contributions
in the two sum rules implies that a simultaneous fit of
both sum rules will not yield their individually optimal
parameter values. Indeed, the nucleon mass increases ($m
=1.06 {\rm GeV}$) and the fit quality does not improve
as significantly as in the $m$ sum-rule. This might
indicate some missing physics in the $\rlap/{q}$ sum-rule.
Radiative O($\alpha_s$) corrections to the conventional OPE,
{\it e.g.}, are known to contribute at the same order as
the leading perturbative term to the $\rlap/{q}$ sum-rule
and their inclusion lowers the nucleon mass \cite{ovc88}.

As mentioned above, the quality of the sum rules is
reflected in the approximate Borel-mass independence of
$m(M)$ and $\tilde{\lambda}_N^2 (M)$, obtained by solving
the optimized sum rules. The instanton-improved sum rules
leave both quantities quite insensitive to the Borel mass
(Fig. 3). The usually less stable $m$ sum-rule, which
receives the main instanton corrections, shows a particularly
pronounced improvement and generates perfect stability
plateaus over the whole fiducial region.

The extension of our investigation to other baryons and
to finite density and temperature is in progress. This
work was supported in part by the US Department of Energy
under Grant No. DE-FG02-93ER-40762.

\begin{figure}
\caption{a) The OPE (dashed line) and instanton
(dot-dashed line) contributions to the $\rlap/{q}$
sum rule. Their sum (dotted line) is compared to the
pole contributions (solid line). b) The same for the
$m$ sum rule. The sum of OPE and instanton contributions
is practically indistinguishable from the pole term.}
\label{fig1}
\end{figure}

\begin{figure}
\caption{The nucleon mass as a function of the Borel
mass from (a) the $\rlap/{q}$ sum rule, (b) the $m$
sum-rule and (c) the combined fit. The coupling
$ \tilde{\lambda}_N^2$ from (d) the $\rlap/{q}$
sum-rule and (e) the $m$ sum-rule.}
\label{fig2}
\end{figure}


\begin{thebibliography}{99}
\bibitem{fn1} For simplicity we will often use the term
"instanton" generically, i.e. for instanton and/or
anti-instanton.
\bibitem{fn3} They are consistent with the results of
\cite{DubSmil81}, which restrict instanton-generated
power corrections in hadron correlators to only a few
low-dimensional terms (in the absence of radiative
corrections).
\bibitem{fn4} The distinction between exponential and
power behaviour remains valid beyond the saddle-point
approximation. The instanton-generated power terms in
eq. (\ref{piq}), in particular, are calculated exactly.
\bibitem{fn7} We have evaluated the sum rules in a large
range of values for t. Their current dependence, both
in fit quality and results, is reduced by the instanton
contributions, as compared to the standard OPE. Due to
space limitations, a detailed discussion will be
postponed to a forthcoming publication  \cite{FoBan92}.
\end{thebibliography}
\end{document}